\documentclass[slac_one]{revtex4}
\usepackage{graphicx}
\usepackage{fancyhdr}
\begin{document}

\title{Measurements of \boldmath $B$ \unboldmath hadron lifetimes at CDF} 

%

\author{S. Behari (for the CDF collaboration)}
\affiliation{The Johns Hopkins University, Baltimore, MD 21218, USA}

\begin{abstract}
Using data samples in excess of 1 fb$^{-1}$ collected by the CDF II
detector we present several world's best measurements of $B$ hadron 
lifetimes. They include $B_s$ meson lifetime in a flavor-specific decay
mode, combining fully and partially reconstructed hadronic decays, 
$B_c^+$ 
meson lifetimes in semileptonic $B_c^+ \rightarrow J/\psi \ell^+ X, 
\ell^+=\mu^+,e^+$ decays and $\Lambda_b$ baryon lifetime in 
$\Lambda_c^+ \pi^-$ 
fully-reconstructed decays. In addition, we introduce a Monte Carlo 
independent technique for measuring $B$ hadron lifetimes in data 
samples biased by displaced vertex triggers.
\end{abstract}

\maketitle

\thispagestyle{fancy}


\section{INTRODUCTION} 
In a simple quark spectator model, the lifetimes of
all $B$ hadrons are equal, independent of the flavor of the
light quark(s) bound to the $b$ quark. However, significant
non-spectator effects enter in the third order,
$(\Lambda_{QCD}/m_{b})^{3}$, and higher order terms in the 
Heavy Quark Expansion (HQE)~\cite{bib:hqe} of the decay width, 
leading to the predicted 
lifetime hierarchy: $\tau(B^+) \geq \tau(B^0) \sim \tau(B_s^0) >
\tau(\Lambda_b^0) \gg \tau(B_c^+)$. Especially, Pauli
Interference terms increase $\tau(B^+)$ by 5\% and 
$\tau(\Lambda_b)$ by 3\% and Weak Annihilation and Exchange
reduce $\tau(\Lambda_b)$ by 7\%.
Recent theoretical calculations predict $\tau(B^+)/\tau(B^0)
= 1.06 \pm 0.02$, $\tau(B_s)/\tau(B^0) = 1.00 \pm 0.01$ and
$\tau(\Lambda_b)/\tau(B^0) = 0.88 \pm 0.05$~\cite{bib:theo}. 
The experimental 
world averages for these ratios are $1.071 \pm 0.009$, 
$0.939 \pm 0.021$ and $0.904 \pm 0.032$, respectively~\cite{bib:pdg}.
While $B^+$ and $B^0$ lifetimes are measured precisely at the
$B$-factories, the experimental uncertainty on $B_s$ lifetime is 
far higher than the theory uncertainty. Prior to 2006 world average 
of $\tau(\Lambda_b)/\tau(B^0)$ was small compared to the 3rd 
order HQE prediction, while the value measured by CDF in 
2006~\cite{bib:cdflb06} is precise but larger than previous
results.

The doubly heavy $B_c^+$ meson decay is an interesting laboratory
to probe heavy quark dynamics, where both $b$ and $c$ quarks can
decay weakly as well as annihilate. Due to significant contributions
from $c$ quark decay and annihilation process, the $B_c^+$ lifetime
is expected to be shorter than the light $B$ mesons. Current 
theory predictions for $\tau(B_c^+)$ range from 0.47 to 0.59 
ps~\cite{bib:bcpred}.

In this proceeding we present world best measurements of $B_s$,
$\Lambda_b$ and $B_c^+$ lifetimes by CDF 
which will help constrain the theories. In addition, we present
a measurement of $B^+$ lifetime in a decay time biased data sample
using a Monte Carlo independent method. It significantly reduces 
systematic uncertainty on lifetime compared to the traditional
Monte Carlo based method and can provide improvements on future
measurements of other $B$ hadrons.

\section{\boldmath $B_c^+$ \unboldmath LIFETIME MEASUREMENTS}
\label{BcLife}
The $B_c^+$ lifetime is measured in $B_c^+ \rightarrow J/\psi \ell^+ X$,
$\ell^+ = e^+, \mu^+$ semileptonic decay channels~\cite{bib:bcmeas} 
collected by di-muon 
trigger in 1.0 fb$^{-1}$ data. Due to the missing neutrino
momentum, the observed decay time ($ct^*$) is corrected by a $K$-factor
as follows:
\( ct = K \cdot ct^* \)
where,
\( ct^* = \frac{M(B_c^+) L_{xy}(J/\psi \ell^+)}{p_T(J/\psi \ell^+)} \)
and
\( K = p_T(J/\psi \ell^+) / p_T(B_c^+) \),
extracted from Monte Carlo.

About 5.6 million signal $J/\psi$ events were reconstructed. The main
challenge in the analysis comes from various backgrounds which enter
due to using a broad partially reconstructed mass shape. These include 
real $J/\psi$ plus misidentified lepton, random track pair consistent
to be $J/\psi$ plus real lepton,
real $J/\psi$ plus real lepton coming from different $b$ quarks, prompt
$J/\psi$ plus lepton and residual conversion in $J/\psi e^+$ decays. The
decay time distribution shape of all backgrounds are modeled and calibrated
carefully both from data and Monte Carlo. The lifetime is extracted
from an unbinned maximum likelihood fit where the signal and background
normalizations are fixed and $\tau(B_c^+)$ is the only free parameter.

The measured lifetimes from the electron and muon channels are
\( c\tau = 121.7 ^{+18.0}_{-16.3} ~\mu\rm{m} \)
and
\( c\tau = 179.1 ^{+32.6}_{-27.2} ~\mu\rm{m} \)
and the combined result is
\( \tau(B_c^+) = 0.475 ^{+0.052}_{-0.049} {\rm(stat)}
                     \pm 0.018 {\rm(syst)} \rm{ps} \)
which is the world best measurement to date.
%
%
\begin{figure*}[t]
\includegraphics[width=70mm]{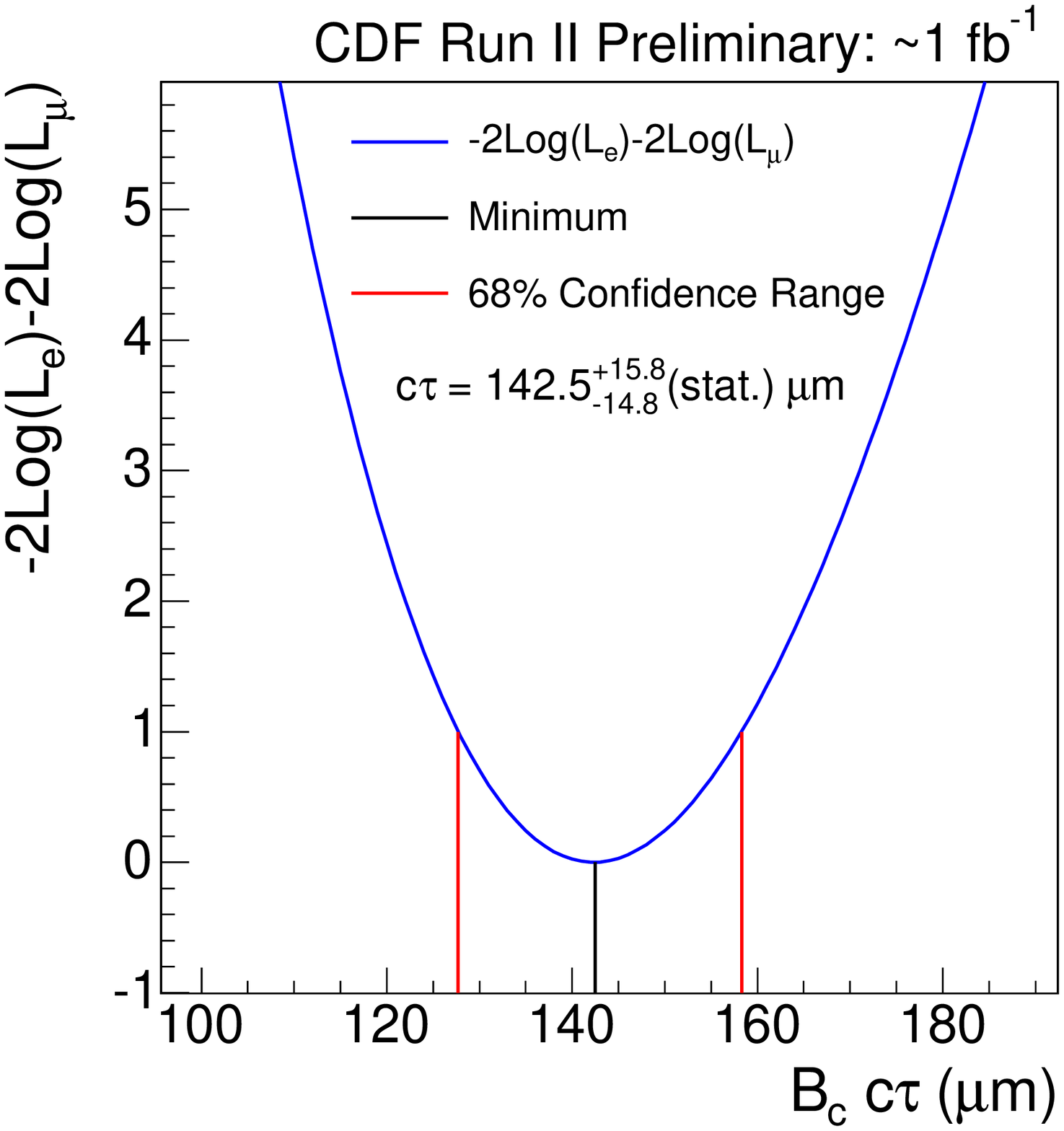}%
\hspace*{0.3in}
\includegraphics[width=70mm]{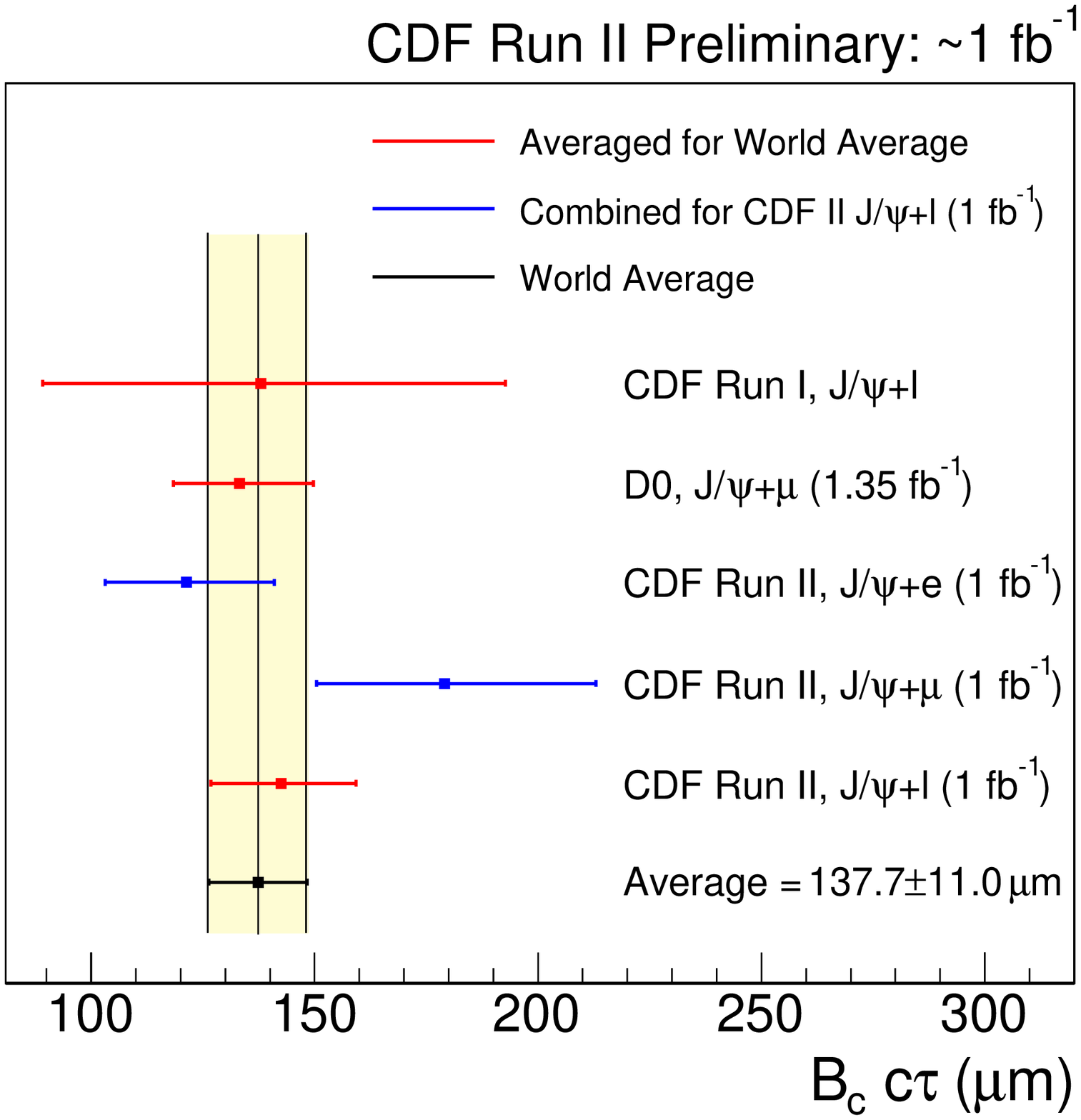}%
\caption{The combined likelihood in muon and electron channels as a 
        function of the $c\tau(B_c^+)$ (left). Comparison of $B_c^+$ 
        lifetime with CDF Run I result and CDF and D\O\ Run II results
        (right). A weighted
        average is made assuming no correlation among these
        measurements. \label{fig:bcctau}}
\end{figure*}
Figure~\ref{fig:bcctau} (left) shows the combined likelihood as a
function of the $c\tau(B_c^+)$. Combining with the D\O\ 
measurements~\cite{bib:pdg} the Tevatron weighted average is
\( \tau = 0.459 \pm 0.037 ~\rm{ps}. \)
%
%
A comparison of our $B_c^+$ lifetime result with Run I CDF and D\O\
results is shown in Figure~\ref{fig:bcctau} (right). A good agreement
is seen between all the measurements.

\section{\boldmath $B_s$ \unboldmath LIFETIME MEASUREMENT}
\label{BsLife}
The $B_s$ lifetime is measured in the flavor-specific 
$B_s \rightarrow D_s^- \pi^+ X$ decays~\cite{bib:bsmeas}, with 
$D_s^- \rightarrow \phi \pi^-$ collected using CDF displaced
vertex trigger in a 1.3 fb$^{-1}$ data sample. In addition to about 
1100 fully reconstructed $D_s^- \pi^+$ events this analysis uses about 2000 
partially reconstructed $D_s^* \pi^+$ and $D_s^- \rho^+(\pi^+ \pi^0)$ events
to increase the statistics significantly. The latter type events involve
missing particles (e.g. a $\pi^0$ in the $D_s^- \rho^+$ mode) which result
in mis-reconstructed mass and transverse momentum. A $K$-factor is 
introduced to correct the lifetime:
\( ct = \frac{L_{xy} \cdot m^{rec}_B}{p_T} \cdot K. \)

The $B_s$ lifetime is determined from two sequential fits. The first is 
a binned maximum likelihood fit of the invariant mass distribution of 
the $D_s^- \pi^+$ candidates to determine the signal and background 
composition. The second is a unbinned maximum likelihood fit of $ct$ 
to extract the $B_s$ lifetime where all the other parameters are fixed. 
Decay time bias due to trigger acceptance is modeled in the final fit by 
an efficiency curve, extracted from Monte Carlo. Extensive tests of the 
fit procedure have been performed on $B^+$ and $B^0$ control samples 
which yield lifetime results in good agreement with the corresponding 
world average values.
%
%
\begin{figure*}
\includegraphics[width=70mm]{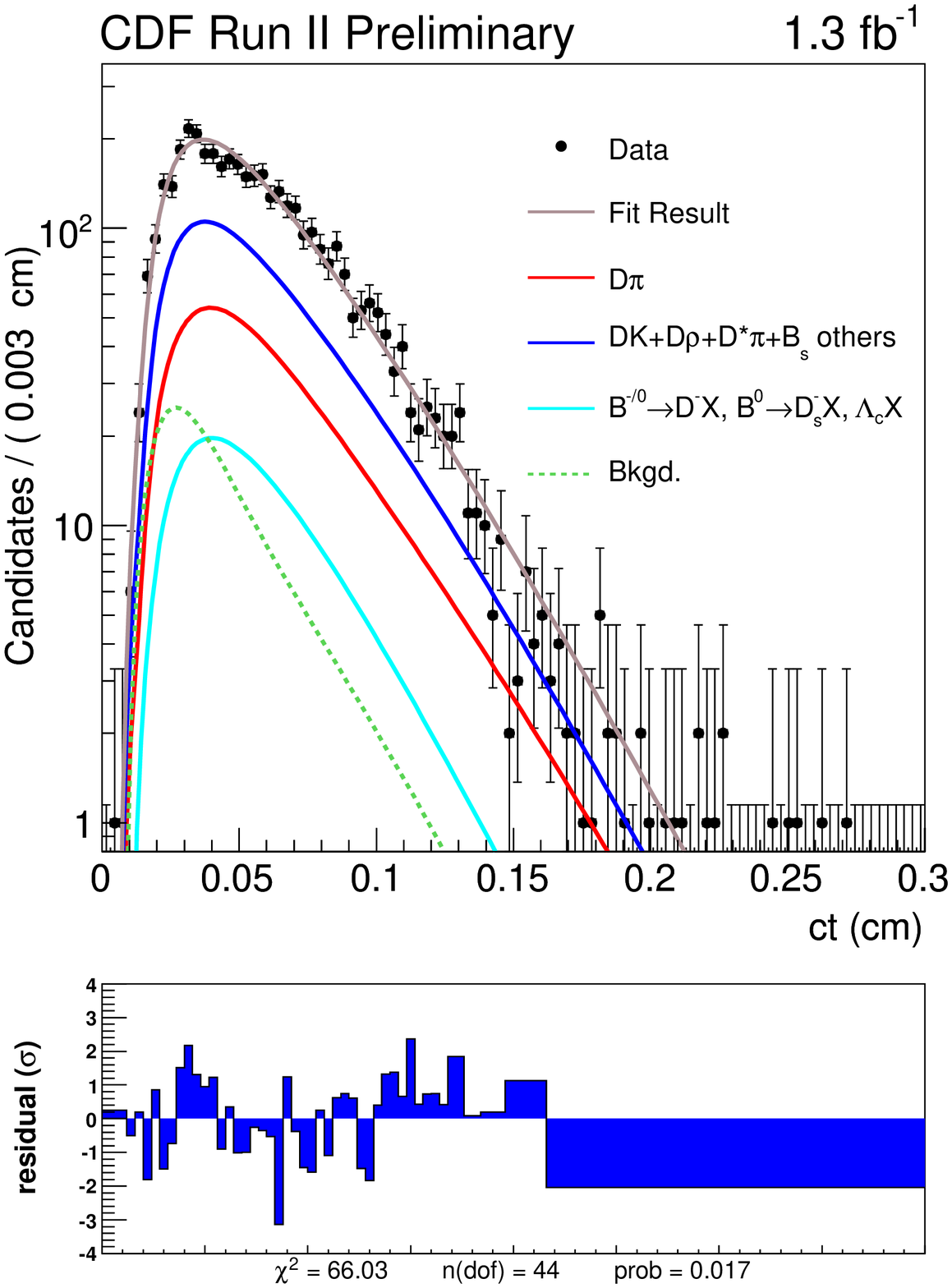}%
\hspace*{0.3in}
\includegraphics[width=80mm]{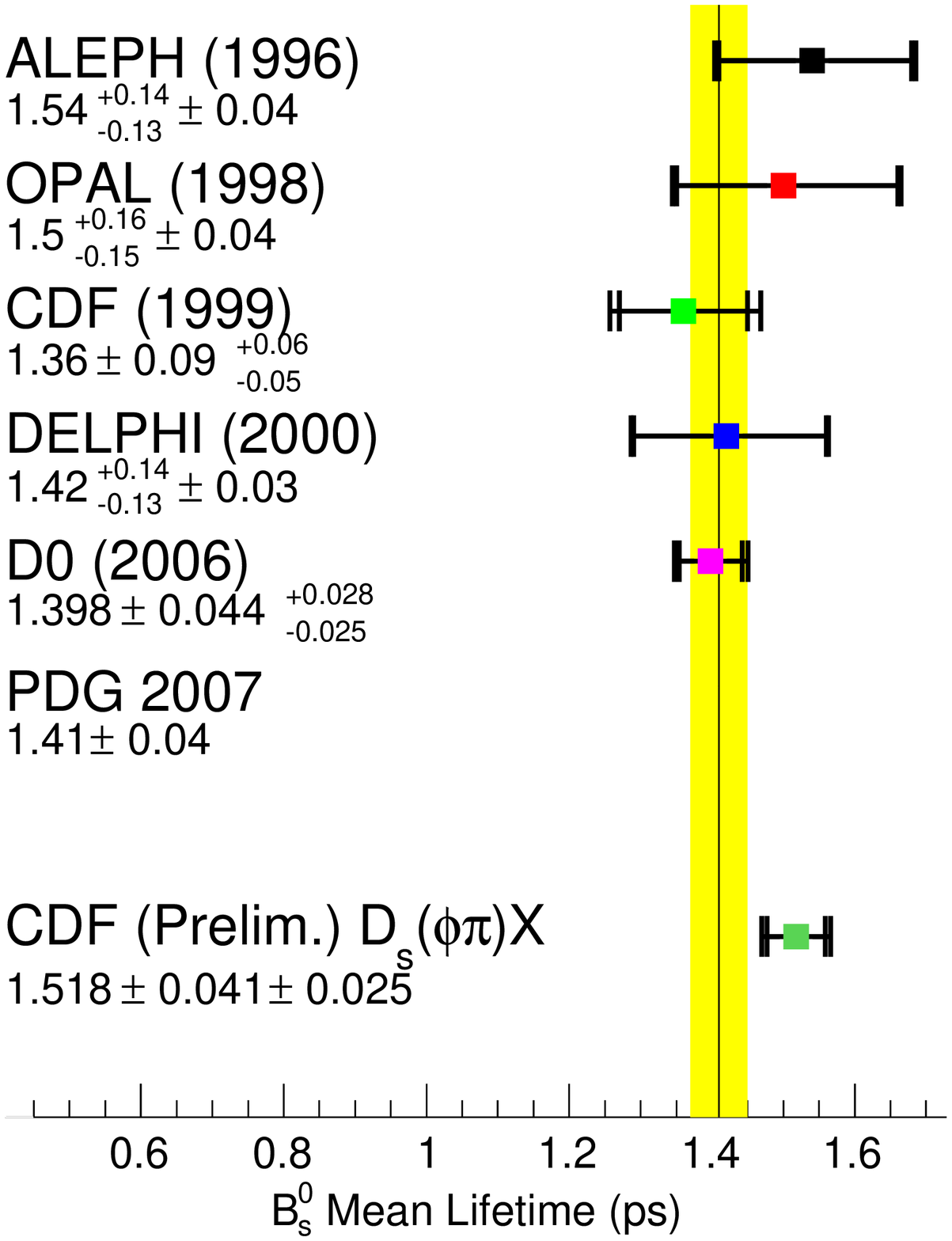}%
\caption{Lifetime fit projection of $B_s \rightarrow D_s^-(\phi\pi^-)\pi^+ X$
        (left). $B_s$ lifetime results from flavor-specific channels
        (right). \label{fig:bsctau}}
\end{figure*}
Figure~\ref{fig:bsctau} (left) shows the lifetime fit projection.
We obtain the flavor-specific $B_s$ lifetime
\( \tau(B_s) = 1.518 \pm 0.041 \rm{(stat)}
                    \pm 0.025 \rm{(syst)} ~\rm{ps}. \)

This is the current world best measurement in flavor-specific mode.
%
%
Figure~\ref{fig:bsctau} (right) compares this result with all 
published results from flavor-specific channels. A good overall
agreement with the PDG 2007~\cite{bib:pdg} result is seen.

\section{\boldmath $\Lambda_b$ \unboldmath LIFETIME MEASUREMENT}
\label{LbLife}
The $\Lambda_b$ lifetime is measured in 
$\Lambda_b \rightarrow \Lambda_c^+ \pi^-$ fully reconstructed 
decays~\cite{bib:lbmeas} obtained
from the CDF displaced vertex trigger where $\Lambda_c$ is reconstructed 
from its decay to a proton, kaon and a pion. About 2900 signal events are 
reconstructed from a data sample of 1.1 fb$^{-1}$. 
A blind analysis is performed where all
analysis procedures are finalized with the signal region 
($m(\Lambda_b) \subset \rm{[5.565,5.670]} \rm{GeV}/c^2$ in data blinded.
As in Section~\ref{BsLife} a two-step fit of invariant mass and lifetime is
performed to extract the $\Lambda_b$ lifetime. As before, the trigger
efficiency curve, obtained from Monte Carlo, is introduced in the 
lifetime fit to model decay time bias. The fit procedure is validated using
input lifetime ($c\tau$) values in Monte Carlo between 325 and 500 $\mu$m.

%
%
\begin{figure*}[t]
\includegraphics[width=70mm]{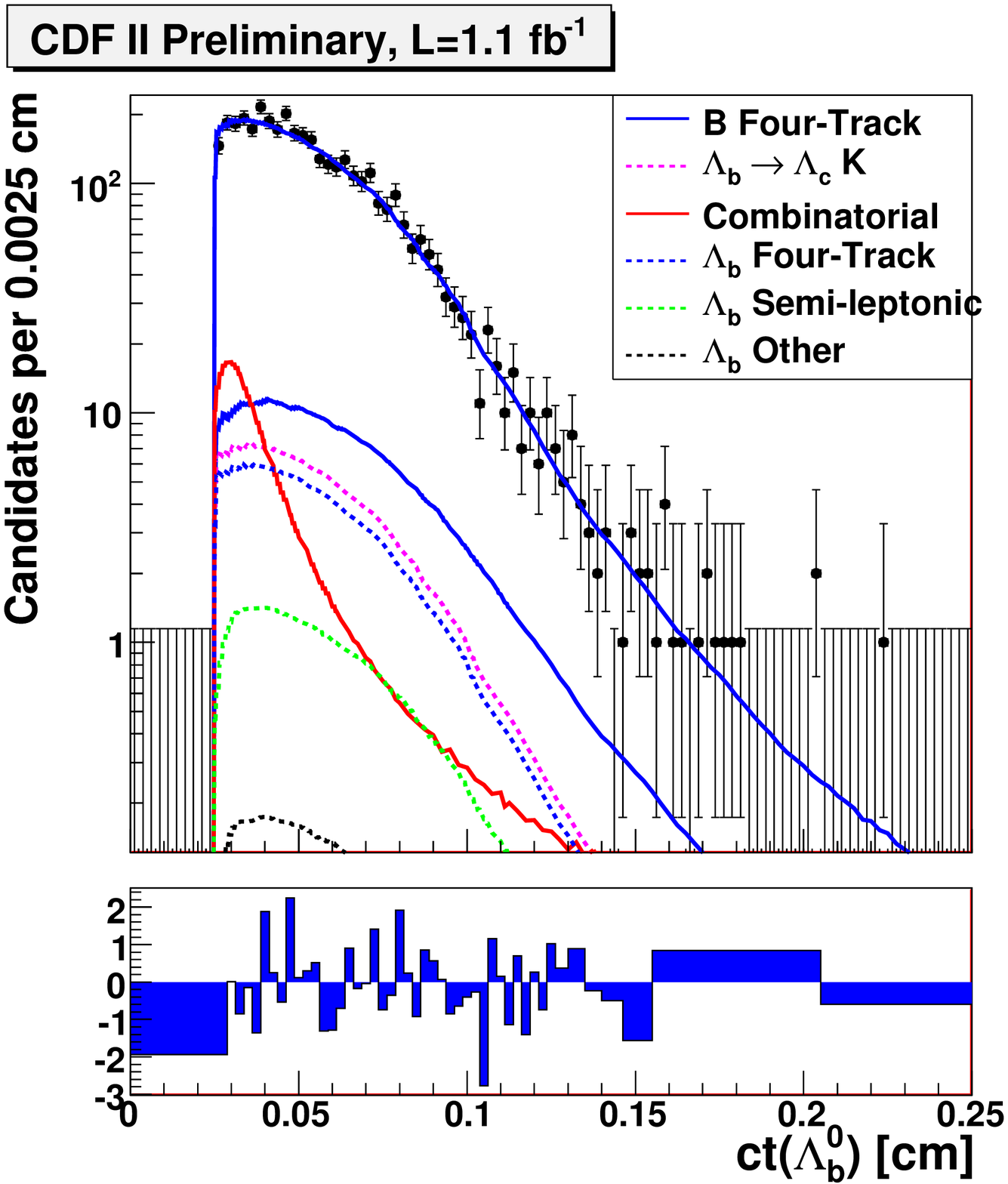}%
\hspace*{0.3in}
\includegraphics[width=85mm]{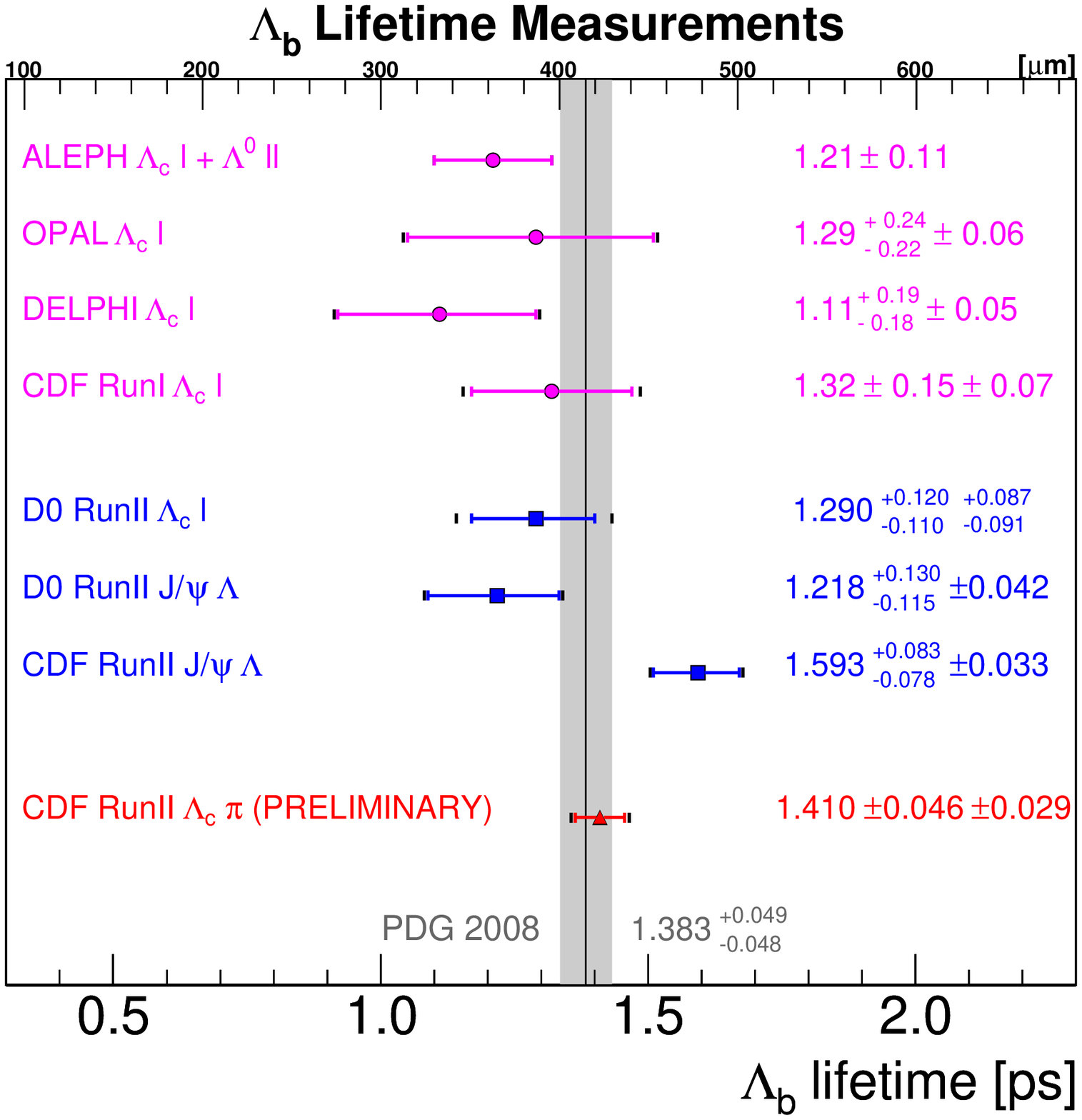}%
\caption{Lifetime fit projection of $\Lambda_b \rightarrow \Lambda_c^+ \pi^-$
       decays (left). $\Lambda_b$ lifetime result compared with PDG 2008
       and other measurements contributing to it (right).
       \label{fig:lbctau}}
\end{figure*}
Figure~\ref{fig:lbctau} (left) shows the lifetime fit projection.
We measure the $\Lambda_b$ lifetime
\( \tau(\Lambda_b) = 1.410 \pm 0.046 \rm{(stat)}
                    \pm 0.029 \rm{(syst)} ~\rm{ps}. \)
This is the current world best measurement. Using world average $B^0$
lifetime~\cite{bib:pdg} we obtain
\( \tau(\Lambda_b)/\tau(B^0) = 0.922 \pm 0.039. \)
The leading sources of systematic error to this measurement come from
modeling of the displaced vertex trigger, $\Lambda_c^+$ Dalitz structure 
and decay time distribution of combinatorial background.
%
%
Figure~\ref{fig:lbctau} (right) compares this result with the current 
world average and all measurements contributing to it, where a good 
agreement is seen. This result is also compatible
with the HQE prediction of the lifetime ratio, $\tau(\Lambda_b)/\tau(B^0)$.

\section{\boldmath $B^+$ \unboldmath LIFETIME MEASUREMENT USING A MONTE CARLO INDEPENDENT METHOD}
\label{BuLife}
The CDF displaced vertex trigger provides heavy-flavor enriched data
sample crucial for many $B$ physics analyses, two of which are presented 
in Sections~\ref{BsLife} and \ref{LbLife}. The decay time distributions 
from this sample are biased which $B_s$ and $\Lambda_b$ lifetime analyses 
correct for, employing Monte Carlo based efficiency functions. This method, 
however, serves as the dominant source of systematic error in both these 
analyses.

A Monte Carlo independent method has been developed which exploits the
decay kinematics in data to correct for the trigger bias on a per event
basis. It completely avoids Monte Carlo modeling of the trigger
effect and thus promises to improve uncertainty involved in future
lifetime measurements in trigger biased samples.

As a proof of principle, we present a measurement of $B^+$ lifetime
measured in the fully reconstructed $B^+ \rightarrow \bar{D^0} (K^+\pi^-) 
\pi^+$ decays~\cite{bib:bumeas}.
About 24200 signal events are reconstructed in 1 fb$^{-1}$ data.
The CDF displaced vertex trigger requires a pair of tracks with large
impact parameters. The Monte Carlo independent method ``slides'' the 
events along the $B$ meson flight direction and evaluates trigger 
acceptance as a function of proper decay length.
%
%
\begin{figure}
\includegraphics[width=85mm]{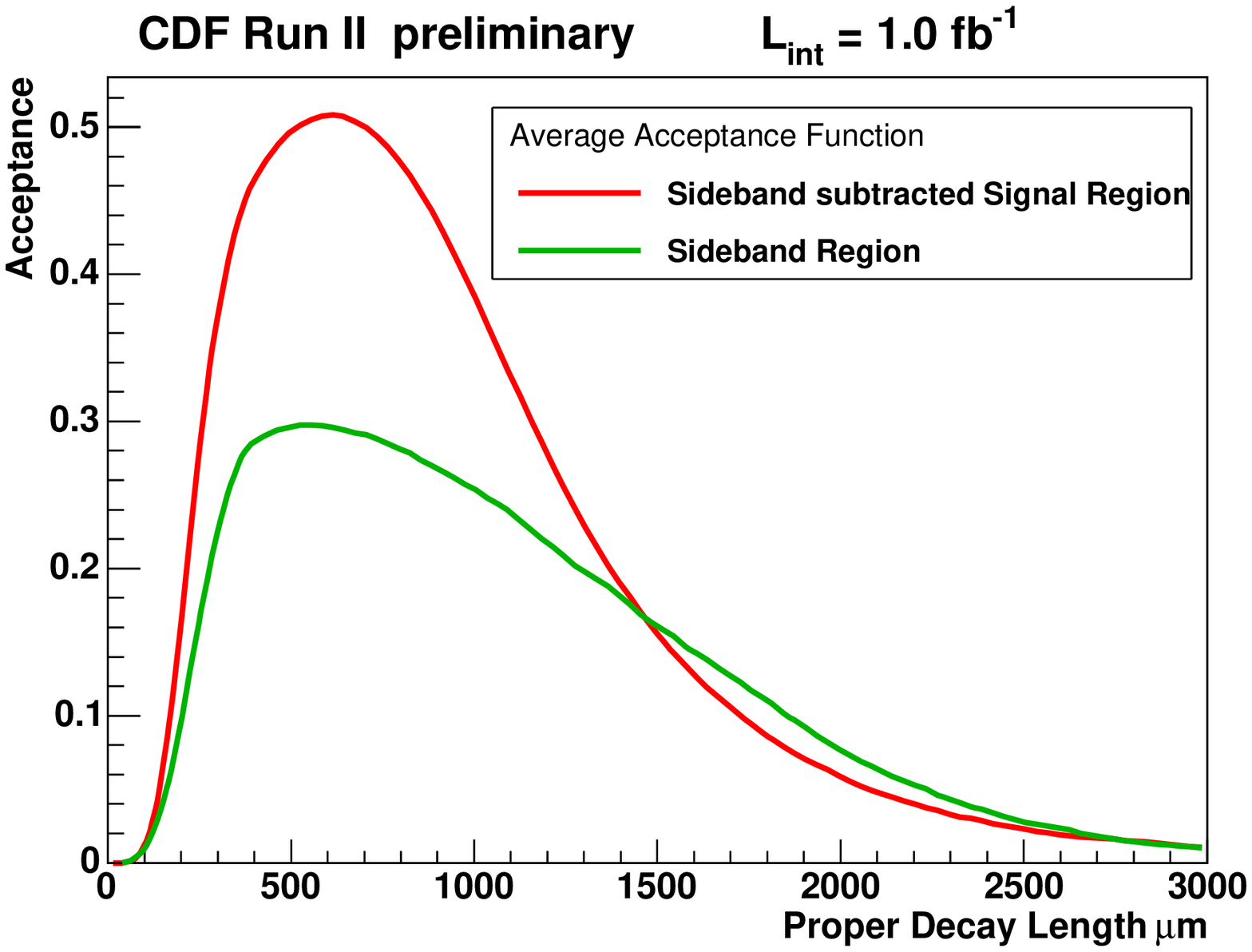}%
\hspace*{0.3in}
\includegraphics[width=70mm]{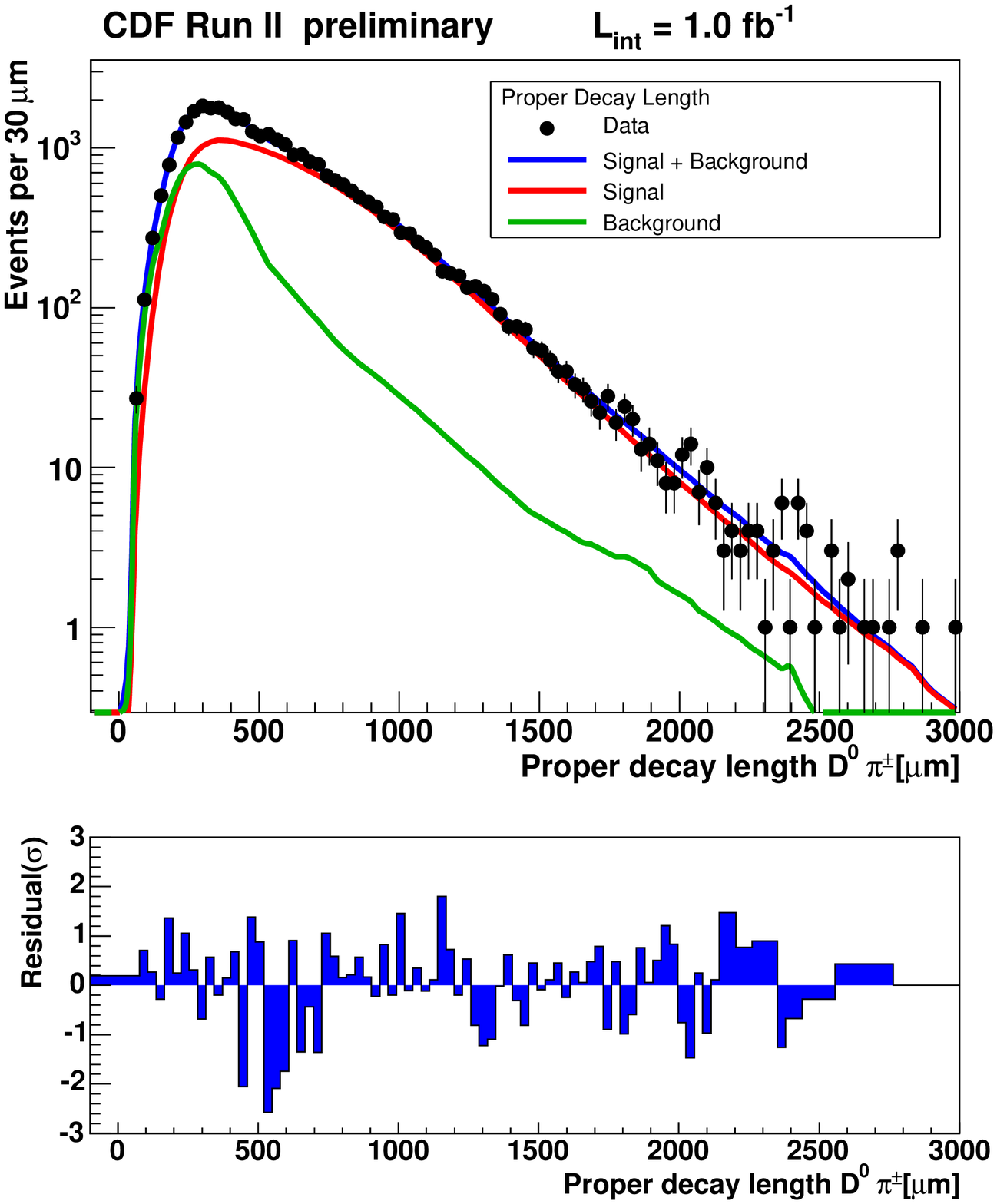}%
\caption{Comparison of averaged acceptance functions for signal and
       background data (left). Lifetime fit projection of
       $B^+ \rightarrow \bar{D^0} \pi^+$ decays (right).
       \label{fig:buctau}}
\end{figure}
The resulting averaged trigger acceptance functions for signal and 
sideband regions are shown in Figure~\ref{fig:buctau} (left). Since the 
per event acceptance is a function and not a scalar variable it greatly 
complicates the fit. To circumvent this, a Fisher Discriminant analysis 
is performed to transform the acceptance into a variable which is easier 
to handle. The following three step method is adopted to extract the 
$B^+$ lifetime. Firstly the mass 
distribution is fitted which defines the signal and sideband regions 
and provides the sample composition. Secondly, using 
the signal and sideband regions the Fisher variable is computed from 
the trigger acceptance. Finally a maximum likelihood fit is performed
on the lifetime and the signal dependent Fisher variable to extract
the lifetime.

%
%
Figure~\ref{fig:buctau} (right) shows the lifetime fit projection.
We measure the $B^+$ lifetime
\( \tau(B^+) = 1.662 \pm 0.023 \rm{(stat)}
                    \pm 0.015 \rm{(syst)} ~\rm{ps}. \)
This is in good agreement with the current world average of 
$1.638 \pm 0.011 ~\rm{ps}$~\cite{bib:pdg}. The small systematic
error of this measurement makes the Monte Carlo independent method
promising for future lifetime measurements of, e.g. $B_s$ and $\Lambda_b$
hadrons.
The dominant sources of systematic uncertainties are
impact parameter and transverse momentum dependence of track finding
efficiency and correlation betwen mass and decay time in the background.

\section{SUMMARY}
Using data samples in excess of 1 fb$^{-1}$ collected by the CDF 
detector we present world best measurements of $B_c^+$,
$B_s$ and $\Lambda_b$ hadron lifetimes. They are in good agreement
with the corresponding experimental world averages and HQE predictions. 
A Monte Carlo independent method is used to measure the $B^+$ lifetime
in a trigger biased data sample which agrees with the current world
average and is expected to be used in future measurements of other
$B$ hadron lifetimes. It is to be noted that all the measurements 
presented in this proceeding are from about 1/4$^{th}$ of the current 
CDF dataset. We expect significant improvements to these results in 
the near future as more data are analyzed.

%


\begin{thebibliography}{9}   

\bibitem{bib:hqe}
I.~I.~Y. Bigi, M.~A.~Schifman and N.~Uraltsev,
Ann.~Rev.~Nucl.~Part.~Sci. {\bf 47} (1997) 591.

\bibitem{bib:theo}
F.~Gabbiani, A.~I.~Onishchenko and A.~A.~Petrov, Phys. Rev. D 70,
094031 (2004) [arXiv:hep-ph/0407004].

\bibitem{bib:pdg}
W.-M.~Yao, et al. (Particle Data Group), J. Phys. G 33, 1 (2006) and
2007 partial update for the 2008 edition.

\bibitem{bib:cdflb06}
A.~Abulencia et al., Phys. Rev. Lett. 98, 122001 (2007). 

\bibitem{bib:bcpred}
V.~V.~Kiselev, arXiv:hep-ph/0308214v1, Aug 2003. 
Phys. Rev. D. 64, 14003 (2001),
V.~V.~Kiselev et al., arXiv:hep-ph/0002127, Feb 2000,
A.~Yu.~Anisimov et al., Phys. Lett. B 452, 129 (1999).

\bibitem{bib:bcmeas}
CDF Public Note 9294,
http://www-cdf.fnal.gov/physics/new/bottom/080327.blessed-BC\_LT\_SemiLeptonic

\bibitem{bib:bsmeas}
CDF Public Note 9203,
http://www-cdf.fnal.gov/physics/new/bottom/080207.blessed-bs-lifetime

\bibitem{bib:lbmeas}
CDF Public Note 9408,
http://www-cdf.fnal.gov/physics/new/bottom/080703.blessed-lblcpi-ct

\bibitem{bib:bumeas}
CDF Public Note 9370,
http://www-cdf.fnal.gov/physics/new/bottom/080612.blessed-MCfree\_Blifetime

\end{thebibliography}
\end{document}